\documentclass[10pt,aps,amsmath,amssymb,prl,twocolumn,superscriptaddress]{revtex4-1}

\usepackage[colorlinks=true,linkcolor=blue,citecolor=blue]{hyperref}%

\usepackage{amsfonts}
\usepackage{dcolumn}
\usepackage{bm}
\usepackage{tikz}
\usepackage{amsmath,amsfonts,amssymb,times,natbib}
\usepackage{multirow, makecell}
\usepackage[standard]{ntheorem}

\begin{document}
\title{Quantum teleportation of physical qubits into logical code-spaces}

\author{Yi-Han Luo}
\thanks{These two authors contributed equally.}
\author{Ming-Cheng Chen}
\thanks{These two authors contributed equally.}
\affiliation{Hefei National Laboratory for Physical Sciences at Microscale and Department of Modern Physics,
University of Science and Technology of China, Hefei, 230026, China}
\affiliation{CAS Centre for Excellence in Quantum Information and Quantum Physics, Hefei, 230026, China}

\author{Manuel Erhard}
\affiliation{Austrian Academy of Sciences, Institute for Quantum Optics and Quantum Information (IQOQI), \\
Boltzmanngasse 3, A-1090 Vienna, Austria}
\affiliation{Vienna Center for Quantum Science and Technology (VCQ), Faculty of Physics,\\
University of Vienna, A-1090 Vienna, Austria}

\author{Han-Sen Zhong}
\author{Dian Wu}
\author{Hao-Yang Tang}
\author{Qi Zhao}
\author{Xi-Lin Wang}
\affiliation{Hefei National Laboratory for Physical Sciences at Microscale and Department of Modern Physics,
University of Science and Technology of China, Hefei, 230026, China}
\affiliation{CAS Centre for Excellence in Quantum Information and Quantum Physics, Hefei, 230026, China}

\author{Keisuke Fujii}
\affiliation{Graduate School of Engineering Science Division of Advanced Electronics and Optical Science \\
Quantum Computing Group, 1-3 Machikaneyama, Toyonaka, Osaka, 560-8531}

\author{Li Li}
\author{Nai-Le Liu}
\affiliation{Hefei National Laboratory for Physical Sciences at Microscale and Department of Modern Physics,
University of Science and Technology of China, Hefei, 230026, China}
\affiliation{CAS Centre for Excellence in Quantum Information and Quantum Physics, Hefei, 230026, China}

\author{Kae Nemoto}
\author{William J. Munro}
\affiliation{NTT Basic Research Laboratories \& NTT Research Center for Theoretical Quantum Physics, NTT Corporation, 3-1 Morinosato-Wakamiya, Atsugi-shi, Kanagawa, 243-0198, Japan }        
\affiliation{National Institute of Informatics, 2-1-2 Hitotsubashi, Chiyoda-ku, Tokyo 101-8430, Japan}

\author{Chao-Yang Lu}
\affiliation{Hefei National Laboratory for Physical Sciences at Microscale and Department of Modern Physics,
University of Science and Technology of China, Hefei, 230026, China}
\affiliation{CAS Centre for Excellence in Quantum Information and Quantum Physics, Hefei, 230026, China}

\author{Anton Zeilinger}
\affiliation{Austrian Academy of Sciences, Institute for Quantum Optics and Quantum Information (IQOQI), \\
Boltzmanngasse 3, A-1090 Vienna, Austria}
\affiliation{Vienna Center for Quantum Science and Technology (VCQ), Faculty of Physics,\\
University of Vienna, A-1090 Vienna, Austria}

\author{Jian-Wei Pan}
\affiliation{Hefei National Laboratory for Physical Sciences at Microscale and Department of Modern Physics,
University of Science and Technology of China, Hefei, 230026, China}
\affiliation{CAS Centre for Excellence in Quantum Information and Quantum Physics, Hefei, 230026, China}

\begin{abstract}
Quantum error correction is an essential tool for reliably performing tasks for processing quantum information on a large scale. However, integration into quantum circuits to achieve these tasks is problematic when one realizes that non-transverse operations, which are essential for universal quantum computation, lead to the spread of errors. Quantum gate teleportation has been proposed as an elegant solution for this. Here, one replaces these fragile, non-transverse inline gates with the generation of specific, highly entangled offline resource states that can be teleported into the circuit to implement the non-transverse gate. As the first important step, we create a maximally entangled state between a physical and an error-correctable logical qubit and use it as a teleportation resource. We then demonstrate the teleportation of quantum information encoded on the physical qubit into the error-corrected logical qubit with fidelities up to 0.786. Our scheme can be designed to be fully fault-tolerant so that it can be used in future large-scale quantum technologies.
\end{abstract}

\maketitle
It is well known that quantum mechanics provides a new paradigm for the creation, manipulation and transmission of information in ways that exceed conventional approaches \cite{Dowling2003,Nielsen2011}. These tasks whether they be in computation, communication or metrology are generally represented by some form of quantum circuit. As the size of these circuits increases, noise and imperfections in the fundamental quantum gates used to implement those circuits render it unreliable to perform the tasks one wanted to do \cite{teleportationfaulttolerant2}. The natural solution is quantum error correction schemes which allows one to construct logical qubits resilient to those errors \cite{Shor96,Gottesman1998,Bennett1996,Devitt_2013}. With logical operations one can then undertake large scale quantum information tasks. It is essential that as part of this, one needs to be able to get ``data'' in and out of the processor in a reliable fashion. 

\begin{figure}[!htp]
\setlength{\abovecaptionskip}{-2 mm}
\setlength{\belowcaptionskip}{-5 mm}
 \centering
            \includegraphics[width=0.95\linewidth]{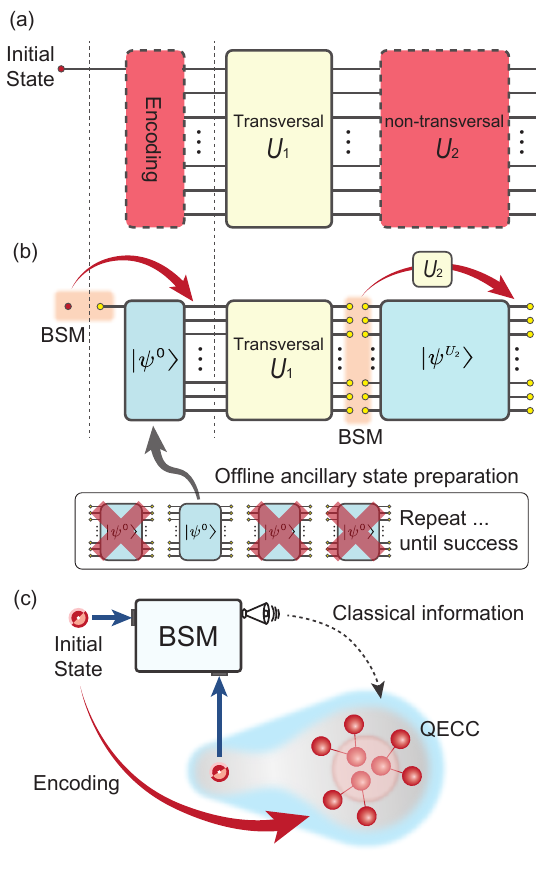}\vspace{3mm}
\caption{Schematic illustration of teleportation based error correction state encoding. In (a) and (b) we show the fault-tolerant quantum circuit before and after combining with quantum teleportation, where the unreliable operations, unknown state encoding and non-transversal gate $U_2$ are marked with red blocks. The flow of quantum information is transmitted along the circuit from left to right. In figure (a), errors will be accumulated as the number of unreliable operations grows. In contrast, by introducing quantum teleportation, the ``fragile nodes'' can be replaced with pre-established entanglement states taking a specific form. As shown in (b), the encoding process and non-transversal gate $U_2$ are replaced with state $|\psi^0\rangle$ and $|\psi^{U_2}\rangle$. Upon encountering ``fragile nodes'', such as encoding, the circuit is paused until a suitable $|\psi^0\rangle = |\Phi^+\rangle$ is generated. Then the BSM transform quantum information holding by the initial state into the QECC, which can then be further operated by following logical gates. Scheme (c) illustrates the teleportation based QECC encoding where to encode the unknown initial state, a physical qubit is entangled with logical qubit encoded in specific QECC. Then the BSM is performed between initial qubit and the physical qubit with the measurement results feedforward to complete the transfer of our quantum information into the QECC. }\vspace{5mm}
\label{fig:scheme}
\end{figure}

Quantum error correction works by encoding the information that is present on a single qubit into a logical qubit, a special type of highly entangled state. This logical qubit has the property that certain errors move the state out of the code space holding the logical qubit \cite{Shor1995}. One can then use ancillary qubits to detect and correct those errors in a non-demolition way \cite{Gottesman1998,Shor1995,Steane1996,Bennett1996,Laflamme1996,Devitt_2013}.  By increasing the redundancy in the degree of freedom within the logical qubit, the errors can be suppressed to arbitrarily low levels.  When the physical error rate is below a certain threshold, it is possible to avoid errors propagating through the circuit to ensure the reliable quantum computation -- a concept known as fault tolerance \cite{Gottesman1998,Shor96,teleportationfaulttolerant2}. It is the key to large scale quantum information processing tasks which generally takes a form illustrated in Fig. \ref{fig:scheme}(a). Here a single qubit holding initial quantum information is encoded into a logical block with the encoding circuit which includes the physical qubits required by quantum error correction code (QECC) and additional ancillary qubits used for the error detection and correction. The encoded logical block is then directed to further logical operation in a fault-tolerant manner. One immediately notices that we have separated these into transversal and non-transversal gates.  The transversal gates have the essential property to prevent errors propagation between physical qubits inside QECC \cite{Eastin2009}. Any QECC requires both transversal and non-transversal gates for universal quantum computation.  Typically most of Clifford gates are transversal and their fault-tolerant implementation is straightforward, whereas non-Clifford gates such as T ($\pi/8$) gate are non-transversal and hence the realization of a logical T ($\pi/8$) gate is  the key for universal quantum computation.

Through introduction of quantum teleportation \cite{teleportation}, these difficulties with non-transversal gates can be addressed. Here we employ a maximally entangled Bell state of the form
\begin{equation}
	|\Phi^+\rangle = \frac{1}{\sqrt{2}}(|0\rangle|0\rangle_\mathrm{L}+|1\rangle|1\rangle_\mathrm{L}), 
	\label{eqn:entanglement}
\end{equation}
where the subscript $\mathrm{L}$ denotes the logical QECC protected state space. As shown in Fig. 1(c), the teleportation utilises a Bell state measurement (BSM) between the initial state $|\psi\rangle$ to be teleported and the single physical qubit of $|\Phi^+\rangle$.  Classical feedforward of our BSM result ensures the initial quantum state to be teleported into the encoded qubit. All these procedures, including the generation of $|\Phi^+\rangle$ together with BSM, can be performed in a  fault-tolerant manner \cite{Nielsen2011}.   
Quantum teleportation allows us to perform non-transversal gates offline, where the probabilistic gate preparation can be done, as shown in Fig. 1(b). The initial state $|\psi\rangle$ could be an arbitrary state, however the choice of the state $|A\rangle = (|0\rangle + e^{i\pi/4}|1\rangle)/\sqrt{2}$, known as a magic state, is the most relevant to quantum computation. It is used to implement the $T$ gate through magic state injection \cite{teleportationfaulttolerant2,Fowler2012} -- a crucial approach towards fault-tolerant non-Clifford gate.  
The same mechanism holds for a fault-tolerant implementation of non-transversal gates when the offline state preparation achieves the required precision though repeat until success strategies. More generally, a recursive application of this protocol allows us to implement a certain class of gates fault-tolerantly, including Toffoli gate \cite{Gottesman1999nature}, which is also indicated in Fig. 1(b).  
It is equally important to note that the quantum teleportation to the logical qubit is an important building block for distributed quantum computation and global quantum communications.
The teleportation based quantum error correction schemes thus have the potential to significantly lower the technical barriers in our pursuit of larger scale quantum information processing.

In stark contrast to theoretical progress,  quantum teleportation and QECC have been developed independently in the experimental regime. We have seen quite a number of remarkable quantum teleportation demonstrations \cite{experimental_teleportation,furusawa1998unconditional,longdistance_teleportation,twoqubit_teleportation, multipledof, satellite, HDtele, bao2012quantum,atom_teleportation1,atom_teleportation2, bao2012quantum,atom_teleportation1,atom_teleportation2, solidstatedefect, superconductive} and QECCs experiments \cite{Lu2008,Yao2012,Nigg2014,Chiaverini2004,Schindler2011,Reed2012,Kelly2015,Ofek2016} performed in a number of physical systems. However the experimental combination of these operations, quantum teleportation based quantum error correction is still to be realized. Given it is an essential tool for future larger scale quantum tasks it will be our focus here.


In this work, we report on the first experimental realization of the teleportation of information encoded on a physical qubit into an error protected logical qubit. This is a key step in the development of quantum teleportation based error correction. We begin by establishing an Einstein-Podolsky-Rosen (EPR) channel -- the entangled resource state for a error protected logical qubit. Quantum teleportation involving a physical qubit of the entangled resource state transfers the quantum information encoded in one single qubit into the error protected logical qubit. The quality of the entanglement resource state and the performance of the quantum teleportation are then evaluated.



\vspace{5mm}
\textbf{Experimental Implementation~}
The scheme shown in Fig. 1(b) is conceptually very similar to the original teleportation protocol, however currently is significantly more challenging due to the necessity of creating the entangled resource Eq.~(\ref{eqn:entanglement}) involving a logical encoded qubit - especially when one considers optical implementations. Here our logical qubit basis states
\begin{equation}
    \begin{array}{ll}
    &\displaystyle |0\rangle_\mathrm{L}=\frac{1}{2\sqrt{2}}(|000\rangle+|111\rangle)^{\otimes3},\\[10pt]
    &\displaystyle |1\rangle_\mathrm{L}=\frac{1}{2\sqrt{2}}(|000\rangle-|111\rangle)^{\otimes3}.
    \end{array}
\label{eqn:shorcode}
\end{equation}
are associated with the (9,1,3) Shor-code \cite{Nielsen2011}, which is a repetition of GHZ$_3$ state \cite{Greenberger1990}. More details concerning Shor code can be found in the supplementary material. Now given the complexity here, it is crucial to design and configure our optical circuit efficiently remembering that in linear optical systems most multiple qubit gates are probabilistic (but heralded) in nature. Only gates including the CNOT gate between different degrees of freedom (DoFs) on the same single photon can be implemented in a deterministic fashion. 

\begin{figure*}[!htp]
    \setlength{\abovecaptionskip}{-2 mm}
    \setlength{\belowcaptionskip}{-5 mm}
        \centering
            \includegraphics[width=.6\linewidth]{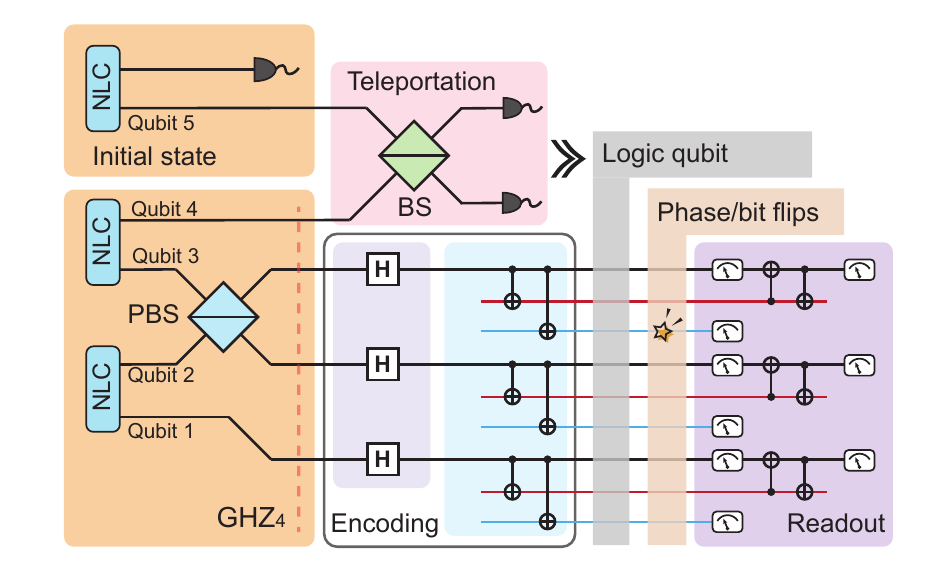}\vspace{3mm}
    \caption{
    \textbf{Experimental setup.} We employ three non-linear crystals (NLC) to create 6 photons in total. Two NLC's in combination with a polarizing beam splitter (PBS) create a four-photon GHZ state in the polarization degree of freedom (DoF). The fifth photon is programmed with an arbitrary qubit state $|\psi\rangle$ to be teleported while the sixth photon serves as a trigger.  Shown in the green box is a beam-splitter (BS) in combination with coincidence detection to implement the Bell-state-measurement (BSM) necessary to teleport the quantum state  $|\psi\rangle$ of the fifth photon into the QECC space. The readout stage (purple box) used to measure the error-syndromes contains three consecutive measurement stages. First, the path DoF is measured followed by the polarization DoF.  Finally, the OAM DoF is measured using a OAM-to-polarization converter. This in total results in eight single-photon detectors (SPD) per photon, thus 24 SPDs for the logic-qubit readout stage only.}\vspace{5mm}
 \label{fig:scheme2}
\end{figure*}

Our experiment is divided into three key stages:
\begin{enumerate}
    \item The creation of the entangled resource state $|\Phi^+\rangle$;
    \item The preparation and teleportation of the initial physical qubit $|\varphi\rangle$ into the logical qubit $|\varphi\rangle_L$;
    \item Readout of the logical state $|\varphi\rangle_L$ and detection of error syndromes.
\end{enumerate}
The first key stage is the creation of the $|\Phi^+\rangle$ state performed using the quantum circuit shown in Fig. \ref{fig:scheme2}(a). It begins by generating a polarization-entangled four photon Greenberger-Horne-Zeilinger ($\text{GHZ}_4$) state \cite{Greenberger1990} using beam-like type-II spontaneous parametric down-conversion (SPDC) in a sandwich-like geometry \cite{Wang2016}. This particular geometry produces a maximally entangled two-photon state and so in order to create a four-photon GHZ state, photons 2 and 3 are combined on a polarization beam splitter (PBS), which transmits horizontally ($H$) polarized photons and reflects vertically ($V$) polarized photons. A four-fold coincidence registration projects the four photons into the GHZ state $|\psi^4\rangle=(|H\rangle^{\otimes4}+|V\rangle^{\otimes4})/\sqrt{2}$. Among these four photons, photon 4 acts as the physical qubit to be used in the BSM while photons 1, 2, 3 are directed to the logical qubit encoding circuit. Now to construct the  9-qubit Shor code with three photons, we use two more degrees of freedom (DoF) per photon associated with the path and orbital angular momentum (OAM). Using additional DoFs is not only resource efficient in terms of the number of photons required, but also enables us to use deterministic CNOT gates using linear optical elements only (see supplementary material for details).

Experimentally, the creation of the Shor code (see Fig.~\ref{fig:scheme2}) begins by applying Hadamard gates on the polarization DoF of each photon using a half-wave-plate (HWP) at 22.5 degrees. This transforms the GHZ state to
\begin{equation}
|\psi'^4\rangle=(|H\rangle |+\rangle^{\otimes3}+|V\rangle|-\rangle^{\otimes3})/\sqrt{2},
\end{equation}
where $|\pm\rangle=(|H\rangle\pm|V\rangle)/\sqrt{2}$ denotes the diagonal/anti-diagonal polarization, respectively. The other DoF are initially in their $|0\rangle$ state. Then two consecutive CNOT gates are applied where the polarization always acts as the control and the other two DoFs as the target qubits. With the control qubit $|\pm\rangle$ and target qubits $|0\rangle$ a three qubit GHZ state $|0,0,0\rangle\pm |1,1,1\rangle$ is generated on each photon. We have thus generated the desired 10-qubit physical -- logical QECC entangled state $|\Phi^+\rangle=(|H\rangle|0\rangle_\mathrm{L}+|V\rangle|1\rangle_\mathrm{L})/\sqrt{2}$ ending the first stage. 

The second stage of the experiment concerns the teleportation of the state $|\varphi\rangle$ on its own independent physical qubit into the QECC protected logical qubit, as depicted in Fig.\ref{fig:scheme}(b). Here we use a photon (photon 5) prepared in a separate BBO crystal (heralded by the second photon of the pair) to encode an arbitrary single qubit state into the polarisation DoF using half and quarter wave plates. A BSM to implement the teleportation is carried out with a 50/50 beam splitter and subsequent coincidence measurement on that polarization encoded qubit and the physical qubit from the entangled resource $ |\Phi^+\rangle$.  Usually, this method projects the two photons onto the anti-symmetric Bell state $\psi^-$, however by transforming the state before the beam splitter using HWPs we project onto the symmetrical $(|HH\rangle+|VV\rangle)/\sqrt{2}$ state \cite{panrmp}.

The third and final stage of the experiment consists of the readout of the encoded qubit. Ideally one should use ancilla qubits to measure the error syndromes and use those results to correct any errors that have occurred before measuring the state of the logical qubit. This of course would require extra photons. However in this case as we want to measure the logical qubit we can independently measure and read out each DoF for photons 1, 2 and 3 without disturbing or destroying the quantum information encoded in the other DoFs \cite{Wang2018}. In our experiment the DoF of polarization, paths and OAM are measured step by step. The qubit encoded with polarization and paths are directly read out with standard polarization analyzers and Mach-Zehnder interferometers respectively while for the OAM encoded qubit a swap gate used to transfer the OAM state to a polarization one where it can be measured with another polarization analyzer. These measurements give us access to the complete logical qubit, consisting of three photons in three different DoFs, and thus access the complete Shor code space of 9 physical qubits. Further details are described in the supplementary materials.

\vspace{5mm}
\textbf{Experimental Results~}
%
The crucial ingredient for our experiment is the generation of the maximally entangled quantum state between the physical and logical qubit. It is important to first evaluate the quality of this entangled resource state. Typical quantum state tomography on ten qubits is unfeasible due to the number of measurements involved.  However, the code structure allows us to eliminate this daunting task to evaluate it at a the physical level.  The logical level evaluation perfectly serves our purpose, and so we instead measure the state fidelity and the CHSH inequality to evaluate the entanglement between the logical and physical qubits. The density matrix of $|\Phi^+\rangle$ can be expressed as
\begin{equation}
\rho = \frac{1}{4}(I\otimes I_\mathrm{cs}+X\otimes X_\mathrm{L}^\mathrm{cs}-Y\otimes Y_\mathrm{L}^\mathrm{cs}+Z\otimes Z_\mathrm{L}^\mathrm{cs}).
\label{eqn:fidelity}
\end{equation}
involving the usual Pauli operators for the physical and logical qubit. Measuring the fidelity is equivalent to determining the expectation values of all four observables above requiring $4\times 2^8=1024$ settings in total. Fortunately, the expectation values of the Pauli matrices $I,Z$ can be obtained with equal settings. Further owing to special features of the Shor-code stabilizers the number of settings can be further reduced to $250$ in total (see supplementary materials). For each setting, we record four-fold coincidences for 10 seconds, yielding a coincidence rate of $\sim$150 s$^{-1}$. We establish a fidelity of $F=0.703(2)$ between the ideal state $|\Phi^+\rangle$.  This clearly surpasses the genuine entanglement $0.5$ threshold. However this fidelity F is insufficient to violate a CHSH inequality with $\langle \mathrm{CHSH}\rangle=1.974(3)<2$ experimental determined. Detailed measurement results for the estimation of the fidelity and CHSH inequality are shown in Fig. 3(a,b) respectively.

\begin{figure}[!htp]
    \setlength{\abovecaptionskip}{-2 mm}
    \setlength{\belowcaptionskip}{-5 mm}
        \centering
            \includegraphics[width=1\linewidth]{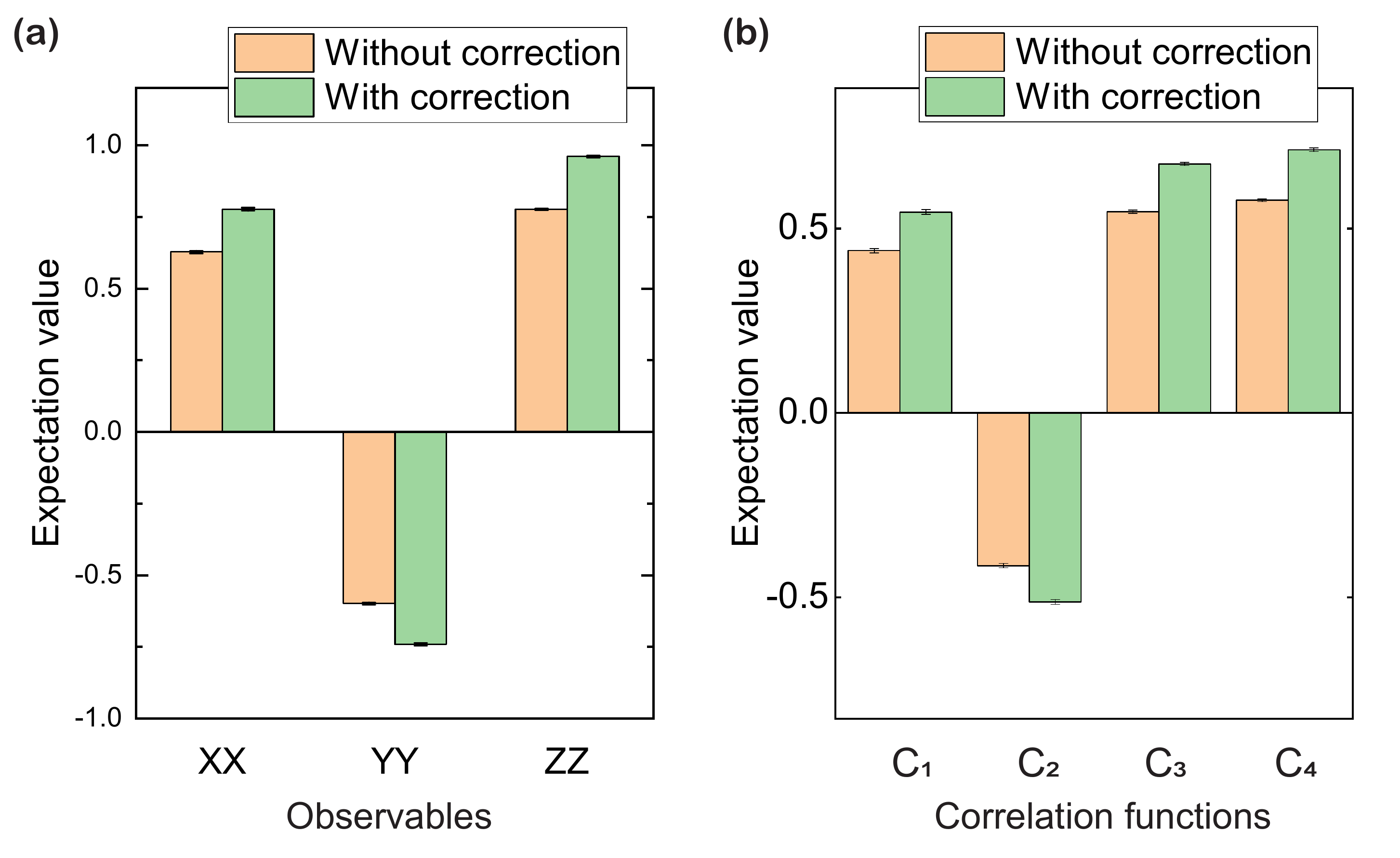}\vspace{3mm}
    \caption{\textbf{Characterization of the entanglement teleportation resource state. } In (a) we show the measured expectation values of $X\otimes X_\mathrm{L}$, $Y\otimes Y_\mathrm{L}$ and $Z\otimes Z_\mathrm{L}$ without (orange bars) and with (green bars) correction. Once can determine the fidelity of entangled state as $F=0.703(2)$ before and $F=0.870(3)$ after correction. Similarly (b) shows the measured correlation functions required for the CHSH inequality without (orange bars) and with (green bars) error correction. The physical qubit is measured in the  $E_{1,2}=(Z\pm X)/\sqrt{2}$ basis while the QEEC is measured with $X_\mathrm{L}, Z_\mathrm{L}$ respectively. The four correlation functions $C_1\sim C_4$ denote $E_1\otimes X_\mathrm{L}$, $E_2\otimes X_\mathrm{L}$, $E_1\otimes Z_\mathrm{L}$ and  $E_2\otimes Z_\mathrm{L}$ respectively. Then $\langle \mathrm{CHSH} \rangle= C_1 -C_2 +C_3+C_4$ gives 1.974(3) before and 2.443(3) after correction. All reported measurements are without background or accidental count subtraction while the stated measurement errors are obtained using Monte Carlo simulation with an underlying Poissonian distribution of photon counting statistics.
    }\vspace{5mm}
    \label{fig:setup2}
\end{figure}

Next, we exclude the influence of correctable errors by confining the state of the logical qubit to the actual code space using the projectors $I_\mathrm{cs}$ to the code space (see supplementary for details). Experimentally, the overlap results in $\langle  I\otimes I_\mathrm{cs}\rangle=0.808(2)$, representing the overlap between the logic qubit prepared in our experiment and the code-space. This is then used to exclude all errors that can be detected by the stabilizers, yielding an error-corrected state fidelity $F=0.870(3)$ and $\langle\mathrm{CHSH}\rangle=2.443(3) > 2$ violation within the code space (see Fig. 3). Furthermore, the encoded state fidelity $F=0.870 > 0.85$ would enable magic state distillation with error-corrected Clifford gates. Our results clearly demonstrate the effectiveness of QECC in our approach but unity fidelity was not achieved due to multi-pair emissions within the SPDC process utilized for generating the $|\Phi^+\rangle$ state. Such errors cannot be corrected by our encoding as they sit inside the code space (see supplementary materials for details). 


With the entangled resource state characterized we now need to explore the operation of teleporting a physical qubit into the logical qubit space. For such a quantum system, it is necessary to show its performance comprehensively exceeding any classical methods. Thus, in our experiment we selected eigenstates with eigenvalue $+1$ of three Pauli matrices $X$, $Y$ and $Z$, denoted as $|0\rangle$, $|+\rangle$ and $|R\rangle$ respectively and measured their teleported fidelity. We measure $125$ settings for $|0\rangle, |R\rangle$ and $98$ settings for $|+\rangle$. For each setting, we accumulate on average $\sim$60 coincidences in 1200 seconds, that corresponds to a count rate of $\sim0.05$ Hz. The achieved experimental fidelities (with and without correction) and the projection probabilities $I_\mathrm{cs}$ are shown in Fig. 4. 
\begin{figure}[!htp]
    \setlength{\abovecaptionskip}{-2 mm}
    \setlength{\belowcaptionskip}{-5 mm}
        \centering
            \includegraphics[width=1\linewidth]{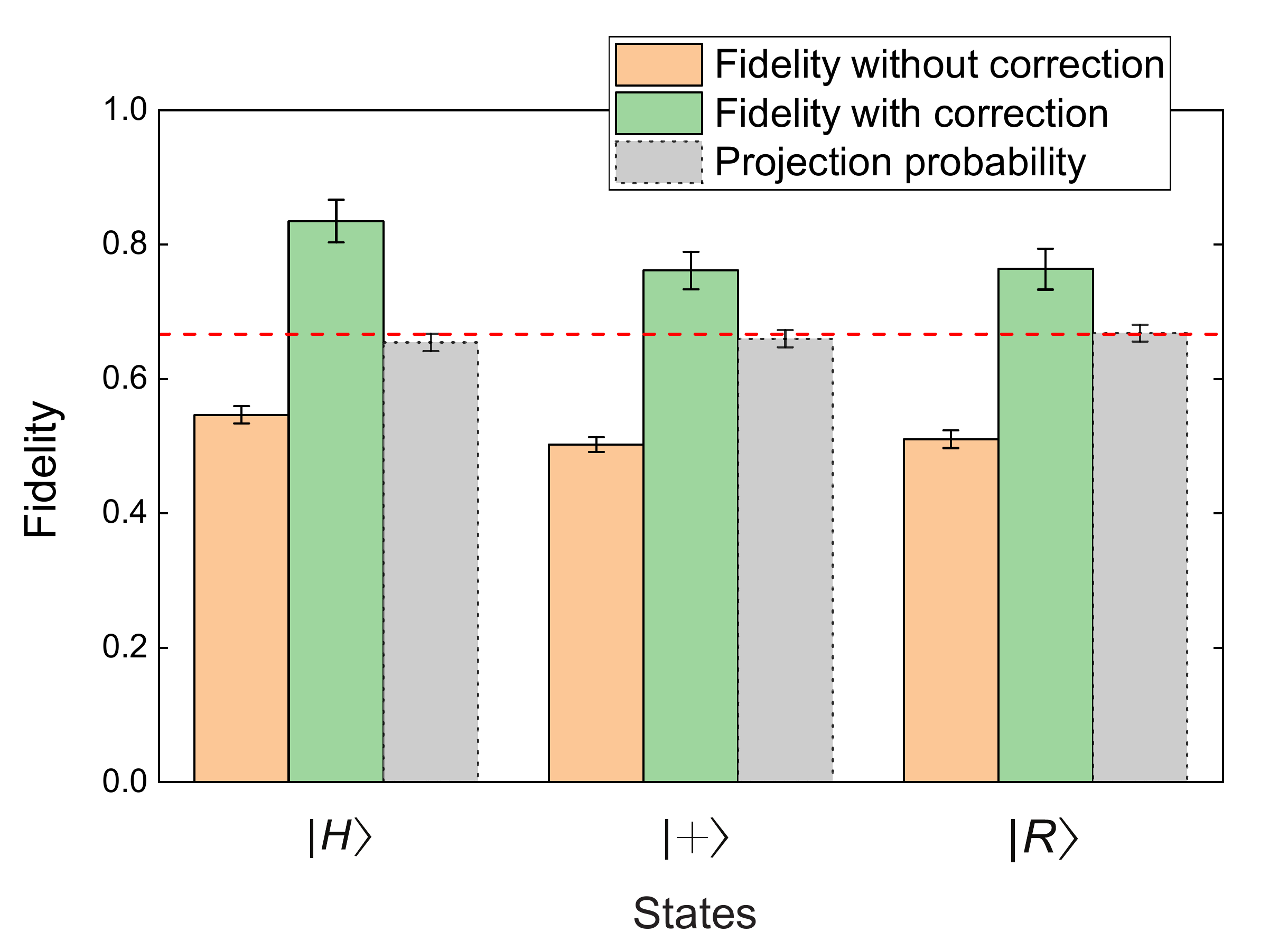}\vspace{3mm}
    \caption{  \textbf{Experimental teleportation of an arbitrary single qubit state.}  Here we show the teleportation results of three representative states $|H\rangle$, $|+\rangle$ and $|R\rangle$ that are eigenstates of $\sigma_{z,x,y}$ respectively with eigenvalue $+1$. For each state the fidelity with and without correction are shown together with the projection probability. After correction the averaged fidelity of the three teleported states is 0.786(17), well exceeding the $2/3$ classical limit shown as a red dashed line.
    }\vspace{5mm}
    \label{fig:setup}
\end{figure}
The averaged fidelity of the three logic states is $0.520(7)$, while after projection into the code space it increases to $0.786(17)$. This is well above the classical limit of $2/3$.  Furthermore, in our experimental arrangements, the teleportation fidelity of any state of the form $(|0\rangle+e^{i\phi}|1\rangle)/\sqrt{2}$ is independent of the phase $\phi$. For example, the fidelities of $\phi=0$ and $\phi=\pi/2$ are consistent in one standard deviation, as shown in Fig. 4.  The obtained results demonstrate the ability of our approach to write via quantum teleportation arbitrary quantum states, including the magic state $\phi =\pi/4$ for $T$-gate, from a single physical qubit into the logical code space consisting of nine physical qubits. Moreover, the post-selected error-correction scheme employed here significantly increases the observed average fidelities from $\sim52\%$ to $\sim78\%$ limited only by non-correctable errors stemming from multi-pair emissions of the SPDC processes.

\vspace{5mm}
\textbf{Discussion and Conclusion~} In summary, we have demonstrated the teleportation of a physical qubit into a logical qubit formed from a QECC. This is a key step for optical quantum calculation on a larger scale. Although the results achieved are far from the fault-tolerance threshold, our work is still far-reaching. It demonstrates the ability to introduce well-developed quantum teleportation to the QIP at the logical level within current technology, and as such represents a crucial step towards fault-tolerant QIP. Such an ability is essential for probabilistic gate operations to be performed on an unknown state in a scalable manner. More specifically and importantly, it allows for magic state injection, a critical task in error-corrected quantum computation. Our experiment can be further modified to adapt the fault-tolerant manner. Moreover, within the theoretical scheme, it can be further concatenated with independently developed modules, such as magic-state distillation and transversal logical operation block, may becomes a useful part of future implementations of fault-tolerant quantum computer or the quantum internet.

\vspace{5mm}

\begin{acknowledgments}
This work was supported by the National Natural Science Foundation of China, the Chinese Academy of Sciences, the National Fundamental Research Program, the Anhui Initiative in Quantum Information Technologies, the MEXT Quantum Leap Flagship Program (MEXT Q-LEAP) Grant No. JPMXS0118069605, the Austrian Federal Ministry of Education, Science and Research (BMBWF) and the University of Vienna via the project QUESS. 
\end{acknowledgments}

%

\end{document}